\documentclass[manuscript,acmsmall,screen]{acmart}
\pdfoutput=1
\usepackage{amsmath,amsfonts}
\usepackage{algorithmic}
\usepackage{graphicx}
\usepackage{textcomp}
\usepackage{xcolor}

\usepackage[ruled,linesnumbered,noend]{algorithm2e}

\usepackage{multirow}
\usepackage{tcolorbox}
 \usepackage{enumitem}
\usepackage{pifont}
 \usepackage[skip=0.5\baselineskip,font=small]{caption}

\usepackage{titlesec}
\usepackage{lipsum}

\titlespacing*{\section}
{0pt}{3pt}{2pt}
\titlespacing*{\subsection}
{0pt}{1pt}{1pt}
\titlespacing*{\subsubsection}
{0pt}{0pt}{5pt}
\AtBeginDocument{%
  \providecommand\BibTeX{{%
    \normalfont B\kern-0.5em{\scshape i\kern-0.25em b}\kern-0.8em\TeX}}}

\newcommand{\cmark}{\ding{51}}%
\newcommand{\xmark}{\ding{55}}%
\newcommand\ours{{\textit{CRIMP}}\xspace}
\begin{document}

\setlength{\abovecaptionskip}{3pt}
\setlength{\belowcaptionskip}{3pt}
\setlength{\textfloatsep}{2pt}
\setlength{\floatsep}{2pt}
\setlength{\dbltextfloatsep}{2pt}
\setlength{\belowdisplayskip}{2pt}
\setlength{\abovedisplayskip}{2pt}

\setlength{\belowdisplayshortskip}{0pt}
\setlength{\abovedisplayshortskip}{0pt}
\title{CRIMP: Compact \& Reliable DNN Inference on In-Memory Processing via Crossbar-Aligned Compression and Non-ideality Adaptation}

\author{Shuo Huai}
\email{shuo001@e.ntu.edu.sg}
		\author{Hao Kong}
\email{kong.hao@ntu.edu.sg}
	\affiliation{
		\institution{School of Computer Science and Engineering \& HP-NTU Digital Manufacturing Corporate Lab, Nanyang Technological University}
			 \city{Singapore}
		 \country{Singapore}
	}
	\author{Xiangzhong Luo}
\email{xiangzho001@e.ntu.edu.sg}

		\author{Shiqing Li}
\email{shiqing.li@ntu.edu.sg}

	\affiliation{
		\institution{School of Computer Science and Engineering, Nanyang Technological University}
			 \city{Singapore}
		 \country{Singapore}
	}

	\author{Ravi Subramaniam}
\email{ravi.subramaniam@hp.com}
	\author{Christian Makaya}
\email{christian.makaya@hp.com}
	\author{Qian Lin}
\email{qian.lin@hp.com}
	\affiliation{
		\institution{HP Inc.}
		\city{Palo Alto}
			 \state{California}
		 \country{United States}
	}
	
	\author{Weichen Liu}
\email{liu@ntu.edu.sg}
\authornote{Corresponding author: Weichen Liu (liu@ntu.edu.sg)}
	\affiliation{
		\institution{School of Computer Science and Engineering, Nanyang Technological University}
		 \country{Singapore}
	}

\begin{abstract}

Crossbar-based In-Memory Processing (IMP) accelerators have been widely adopted to achieve high-speed and low-power computing, alleviating the memory wall issues in the Von Neumann architecture, especially for the deep neural network (DNN) models with numerous parameters and high computational complexity. However, the floating-point (FP) arithmetic is not compatible with crossbar architectures. Although quantization schemes can remove FP parameters, current quantization techniques still require specific FP processors for multiplying scaling factors, incurring large hardware overhead. Besides, redundant parameters of current DNN models occupy too many crossbars, limiting the efficiency of crossbar accelerators. IMP-aware pruning methods are proposed to reduce the number of used crossbars, but current methods require data aligning among crossbars, which introduces significant memory overhead and computing overhead. On the other hand, due to the inherent non-ideal behavior of crossbar devices, like write variations, pre-trained DNN models suffer from accuracy degradation when it is deployed on a crossbar-based IMP accelerator for inference. Although some compensation methods are proposed to diminish the impact of device non-ideality, they introduce much overhead to the hardware design or runtime and do not consider the characteristic of crossbars. Especially, to deploy complex models on IMP accelerators, we should compact the model and mitigate the influence of device non-ideal behaviors without introducing significant overhead from each technology.

In this paper, we first propose to reuse bit-shift units in crossbars for approximately multiplying scaling factors in our quantization scheme to avoid using FP processors. Second, we propose to apply kernel-group pruning and crossbar pruning to eliminate the hardware units for data aligning. Third, we adopt the runtime-aware non-ideality adaptation to relieve the impact of non-ideality from the training stage by exploiting the feature of crossbars. Finally, we integrate these three optimization procedures into one training process to form a comprehensive learning framework for co-optimization, which reduces the training overhead and achieves higher accuracy. The experimental results indicate that our quantization method incurs only a negligible accuracy drop. Our pruning approach achieves a higher sparsity rate and higher accuracy compared to state-of-the-art model compression methods. For our comprehensive learning framework, we obtain significant improvements over the original model on the crossbar-based IMP accelerator, with an average reduction of computing power and computing area by $122.38\times$ and $19.51\times$, respectively. Furthermore, we can obtain totally integer-only, pruned, and reliable VGG-16 and ResNet-56 models for the Cifar-10 dataset on IMP accelerators, with accuracy drops of only $2.19\%$ and $1.26\%$, respectively, without any hardware overhead.

\end{abstract}




\maketitle

\section{Introduction}
\label{section:introduction}

Deep Neural Networks (DNNs) have advanced many fields, including image recognition and natural language processing. Meanwhile, the demand for low-power edge intelligence is rapidly increasing in everyday devices like mobile phones and wearable gadgets \cite{kong2022smart}. However, the efficient deployment of DNNs on low-power devices is hindered by their growing need for computational ability and memory resources. DNN algorithms exhibit a high degree of computing parallelism but require large memory access, thus, the Resistive Random Access Memory (ReRAM)-based In-Memory Processing (IMP) crossbar architecture is an emerging and promising solution for efficiently accelerating these DNN algorithms. ReRAM crossbar architectures exploit parallel digital arithmetic units and perform computations within the memory itself, thereby maximizing parallel execution and minimizing memory access \cite{shafiee2016isaac}. Despite the advantages of IMP crossbar architectures, current DNN models are still challenging to deploy efficiently due to the floating-point (FP) computing and redundant parameters \cite{huai2023crossbar}. Furthermore, the non-ideal behaviors of the resistance of ReRAM cells also pose a significant challenge, reducing the accuracy of pre-trained models during inference on the ReRAM IMP crossbar \cite{klachko2019improving}. To overcome these challenges, some techniques need to be proposed to enable efficient and reliable deployment of DNNs on IMP crossbar architectures.

\begin{figure}[!t]
    \includegraphics[trim={0pt 0pt 0pt 0pt},clip,width=0.99\linewidth]{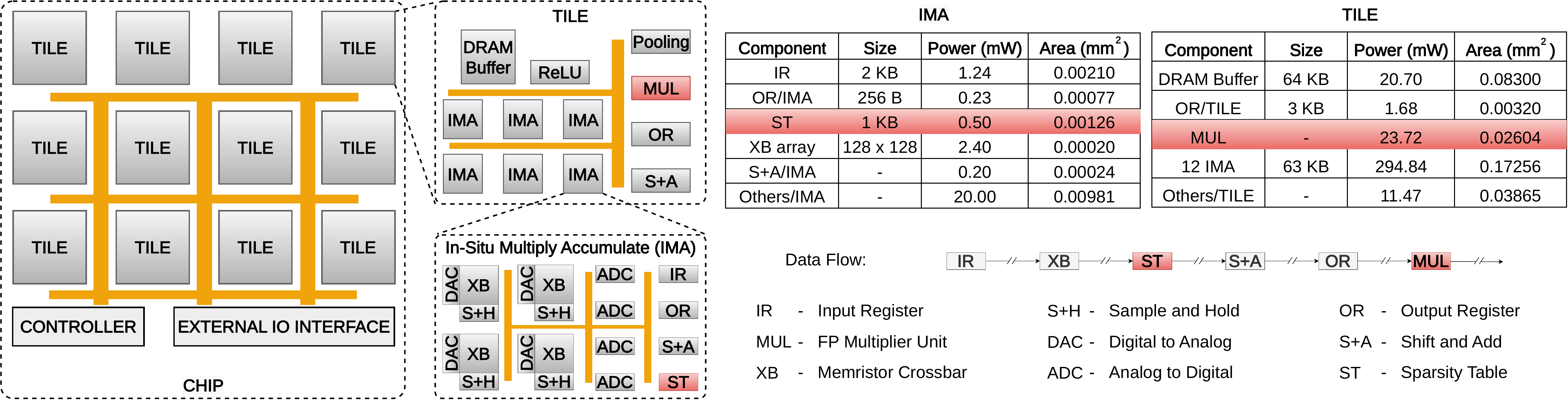} 

\caption{IMP architecture from ISAAC \cite{shafiee2016isaac} with the parameter of each component. Please note that \textit{ST} is used for crossbar-column/row IMP-aware pruning, and \textit{MUL} is used for processing FP multiplications.} \label{fig:cimpara}
\end{figure}

ReRAM crossbar architectures have limited write endurance and high reconfigure latency \cite{yu2016emerging}. It is crucial to minimize the number of writing operations to reduce latency and extend the life of the crossbar. Therefore, during the model inference stage on ReRAM-based IMP architectures, it is impossible to reconfigure the crossbar to infer different layers or parts of a DNN model. But pre-loading the entire model can result in using more crossbars than the IMP device owns. To enable efficient deployment of complex models on IMP devices, the model should be compressed. IMP-aware fine-grained pruning methods \cite{crossbaraware,enablingReRAM, chu2020pim}, which trim columns/rows of weights in each crossbar, can reduce the number of required crossbars. However, these methods necessitate expensive extra hardware to align the output/input of each crossbar \cite{chu2020pim}, which severely decreases the hardware integration density and reduces the number of crossbars that can be embedded on IMP devices, as shown in Fig.~\ref{fig:cimpara}. Thus, new pruning methods should be proposed to effectively deploy  DNN models on IMP devices without introducing additional hardware.

Meanwhile, implementing FP arithmetic units on IMP crossbars is complicated and expensive \cite{ensan2019fpcas}, thus, the crossbar architectures cannot directly perform FP multiplications. This means extra FP processors are required in the IMP device for DNN models with FP operations, which introduce a power overhead of approximately $7$\% and an area overhead of about $9$\% into the IMP device, as shown in Fig.~\ref{fig:cimpara}. To achieve low-power DNNs for IMP devices, the final DNN model should only comprise integer operations. Although traditional model quantization schemes \cite{choi2019accurate} can approximate FP inputs and weights with integers, they still employ FP scaling factors to maintain accuracy. IAO \cite{jacob2018quantization} proposes replacing FP scaling factors with fixed-point multiplication, but this approach still requires a fixed-point multiplier, which also introduces some overhead. 


Furthermore, due to the inherent features of ReRAM cells, it is subject to some non-idealities, such as Write Variation (WV)  \cite{hsu2015study}, Stuck-At-Faults (SAFs) \cite{chen2014rram}, and others. These non-idealities can significantly reduce the accuracy of model inference achieved by ReRAM-based IMP accelerators \cite{charan2020accurate}. One intuitive solution for this issue is the Read-Verify-Write (R-V-W) approach \cite{charan2020accurate}, which continues writing of ReRAM cells until their resistance values converge to the required values. However, this method is associated with an expensive feedback control mechanism and multiple writes to ReRAM, which can lead to significant configuration overheads and decrease the endurance of the ReRAM-based IMP device. Alternatively, some techniques \cite{meng2021digital, he2023rvcomp} try to compensate for the errors caused by ReRAM non-idealities using additional hardware units. Although such approaches do not incur training overhead, the introduction of extra hardware results in significant area and power overheads. Recently, some methods \cite{eldebiky2022correctnet, long2019design} have integrated the non-idealities of ReRAM into model training to mitigate their impact on accuracy without hardware overhead. Nonetheless, these methods do not consider the execution flow of model inference on ReRAM, disregarding the influence of crossbar features such as crossbar size, Analog-to-Digital converters (ADCs), etc.

To address the aforementioned problem, in this paper,  we propose a novel DNN learning framework, named \ours, that can create a \underline{c}ompact and \underline{r}eliable DNN model with high accuracy for \underline{IMP} architecture during one training process. To achieve efficient, low-power, and reliable IMP acceleration, we apply integer-only quantization, crossbar-aligned pruning, and runtime-aware non-ideality adaptation schemes, which are integrated into one training process to reduce the training overhead. Our main contributions are summarized as follows.

%
%

\vspace{-0.1cm}
\begin{itemize}

\item We introduce a crossbar-aligned pruning approach to reduce crossbar usage, without requiring extra processing units for data aligning, thereby improving hardware efficiency without decreasing integration density. We incorporate kernel-group pruning and crossbar pruning to create multi-grained pruning, which improves accuracy and sparsity. Additionally, we apply a simple yet efficient integer-only quantization scheme for the IMP crossbar architecture by reusing the bit-shift units. (see \textbf{Section \ref{subsection:pruning}} and \textbf{Section \ref{subsection:fp}})

\item We adopt a runtime-aware non-ideality adaptation scheme to learn DNN models for ReRAM-based IMP accelerators. In addition, we build a realistic crossbar-based runtime simulator to imitate the actual inference flow on the ReRAM crossbar by considering various non-idealities of ReRAM cells and crossbar features. The simulator models the non-ideality to guide the training stage for error-diminishing. (see \textbf{Section \ref{subsection:variation}})
\item We propose a model learning framework to complete the crossbar-aligned pruning, integer-only quantization, and runtime-aware non-ideality adaptation via one training process, which co-optimizes these techniques to improve accuracy. Additionally, we optimize models with a dynamic zero-recovery procedure, which widens the exploration space for better architecture and better weights, achieving higher accuracy. (see \textbf{Section \ref{subsection:compact}})


\item We demonstrate the effectiveness of \ours with extensive experiments. Specifically, \ours achieves a higher sparsity rate and a slighter accuracy drop without extra hardware, compared to state-of-the-art IMP-aware pruning methods. Moreover, \ours significantly reduces computing power and area, while also producing totally compact and reliable models with only a slight accuracy drop and no overhead.  (see \textbf{Section \ref{section:experiment}})

\end{itemize}
\vspace{-0.1cm}


 The rest of this paper is structured as follows. In Section \ref{section:relatedwork}, we provide an overview of the background and related works on IMP-aware DNN pruning, quantization, and ReRAM non-ideality processing. Section \ref{section:methodology} elaborates on the details of our proposed approaches. Section \ref{section:experiment} presents the experimental results. In Section \ref{section:conclusion}, we conclude this paper.

\section{Background \& Related Works}
\label{section:relatedwork}

\begin{table}[!t]
\centering
\small



\caption{The symbols and their definitions used in the equations or algorithms of this paper.}
\label{tab:symbol}
    \renewcommand{\arraystretch}{1.1} 
\begin{tabular}{ccccc}
\hline
Symbol             & Definition                          & \quad \quad \quad & Symbol        & Definition                        \\ \hline
$r$                & Ideal resistance value              &      & $W$           & The weights of a layer            \\
$r'$               & Actual resistance value             &      & $B$           & The biases of a layer             \\
$\theta$           & Distribution of the error           &      & $S$           & The scaling factor                \\
$e$                & Euler's number                      &      & $Q$           & The quantized value               \\
$\epsilon$         & Standard deviation of the error     &      & $Z$           & The zero point                    \\
$N$                & Total number or Normal distribution &      & $\circledast$ & The convolutional operation       \\
$C$                & The number of channels/kernels      &      & $\partial$           & Partial derivative                         \\
$C'$               & The number of remaining kernels     &      & $D$, $d$      & The dimension (length) of vector  \\
$K$                & The size of kernels                 &      & $\delta$      & The importance factor of weights      \\
$X$                & The input of a layer                &      & $XB_s$        & The size of crossbar              \\
$\gamma$           & The weights in BN layer             &      & $s$           & The epoch to start zero-recovery  \\
$lr$         & Learning rate                  &      & $c$           & Ideal conductance (weight) value                  \\
$XB$               & The abbreviation of crossbar        &      & $c'$          & Actual conductance (weight) value \\
$H$, $h$           & The height of crossbar              &      & $q$           & The number of quantization bits   \\
$w$                & The width of crossbar               &      & $p$           & The pruning ratio                 \\
$L$, $l$           & Layer index in DNN                  &      & $\sigma$      & The activation function           \\
$n$, $a$, $b$, $m$ & Any number                          &      & $z$           & Updated value in a training step  \\
$v$                & Voltage value, each bit of the input                      &      & $t$           & Training step     \\ \hline
\end{tabular}
\end{table}
In this section, we provide a brief on the fundamentals of IMP, as well as an overview of IMP-aware pruning, quantization, and non-ideality processing. Additionally, we discuss the advantages and disadvantages of some related works in this field, establishing motivation for our proposed approach.  Table \ref{tab:symbol} shows the symbols and definitions used in this paper.

 The ReRAM-based IMP device is composed of numerous crossbars, with memristive cells situated between each horizontal wordline and vertical bitline in a crossbar \cite{shafiee2016isaac}. When inference DNNs, the weights of a model are mapped to the memristive cells, and the conductance represents the associated weight values. And the digital-to-analog converter (DAC) converts the input to voltage pulses that are injected into the wordline of the crossbar. According to Kirchoff's Law, the current generated in each cell, representing the product between the voltage and the cell conductance, accumulates along the bitline, and the total current is the dot product result. This result is then converted to digital values by the analog-to-digital converter (ADC). As matrix multiplication can be performed on crossbars easily, fully-connected layers can be directly mapped to crossbars.

\begin{figure}[!t]

    \includegraphics[trim={0pt 0pt 0pt 0pt},clip,width=0.99\linewidth]{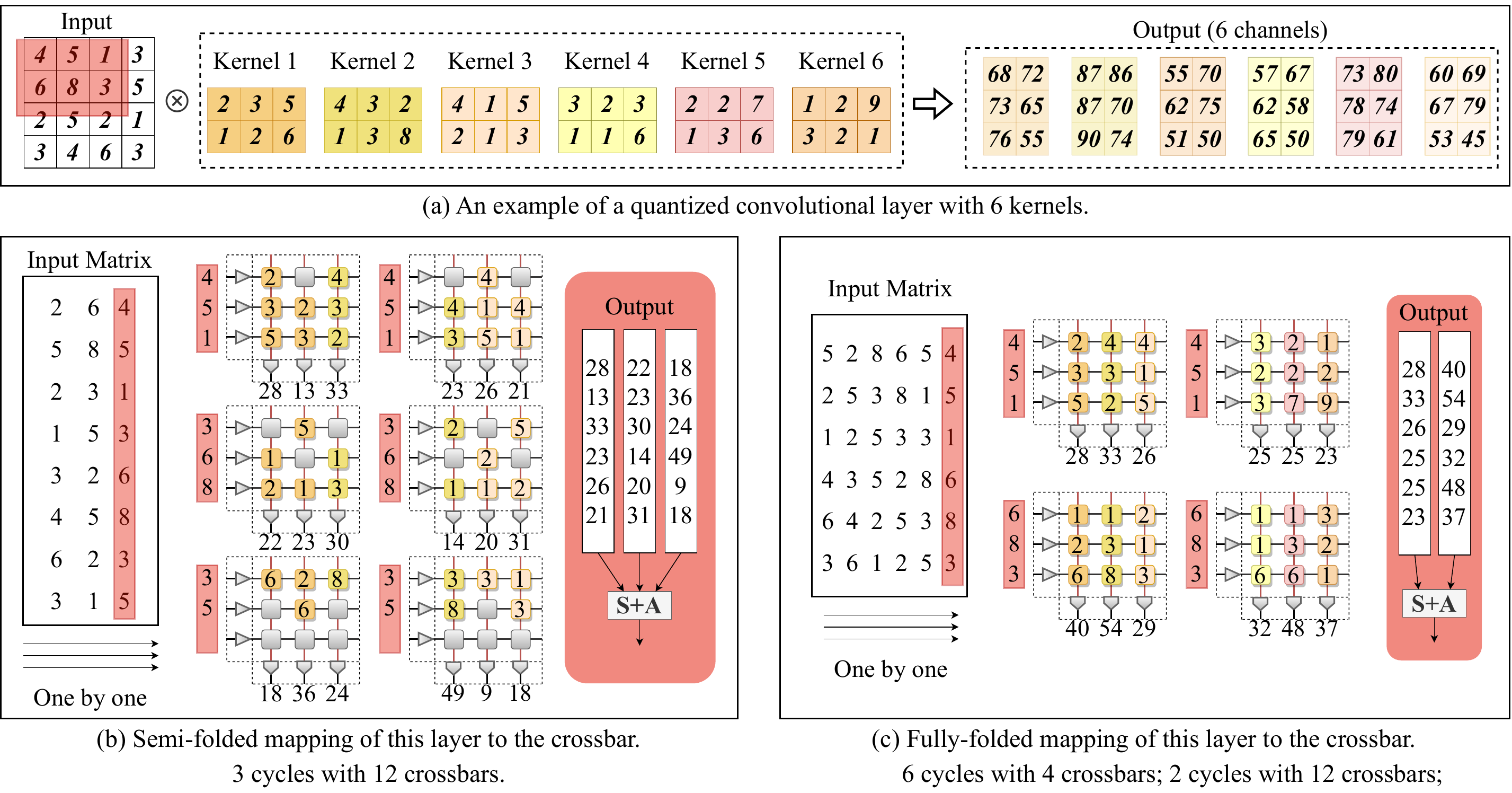} 


\caption{(a) An example of a quantized convolutional layer; (b-c) Different mapping methods for this layer on crossbars, and their execution cycles and the number of used crossbars. 
} \label{fig:mapping}

\end{figure}

To demonstrate the mapping of convolutional layers on crossbars, Fig.~\ref{fig:mapping} (a) shows an example of a quantized convolutional layer and Fig.~\ref{fig:mapping} (b-c) illustrate different mapping methods for this layer on crossbars. In the semi-folded mapping method, as shown in Fig.~\ref{fig:mapping} (b), several whole rows of the input (at least can be multiplied by the kernel) is first unfolded into a vector, and the kernel weights are then mapped to the corresponding crossbar memristive cells. Since the unfolded input vector needs to be multiplied by different parts of the kernel, each kernel must be duplicated several times. On the other hand, the fully-folded mapping method, as depicted in Fig.~\ref{fig:mapping} (c), unfolds each kernel to a vector and maps it to the crossbar memristive cells. The input of each layer is reshaped to a series of vectors with the same shape as the unfolded kernel vectors, and sent to the wordline of crossbars one by one. In both mapping methods, a large kernel occupies several crossbars that are connected by peripheral circuits. However, there are some non-used memristive cells in the semi-folded mapping approach. Conversely, the fully-folded mapping method can fully occupy the crossbar. Additionally, the weights in the fully-folded mapping method can also be duplicated several times to enable different input voltage injections in parallel among different replicas, which increases the parallelism of this method \cite{shafiee2016isaac, chi2016prime} to reduce execution cycles. Thus, our proposed approach focuses on optimizing DNN models based on the fully-folded mapping scheme.

\begin{figure}[!t]

    \includegraphics[trim={0pt 0pt 0pt 0pt},clip,width=0.99\linewidth]{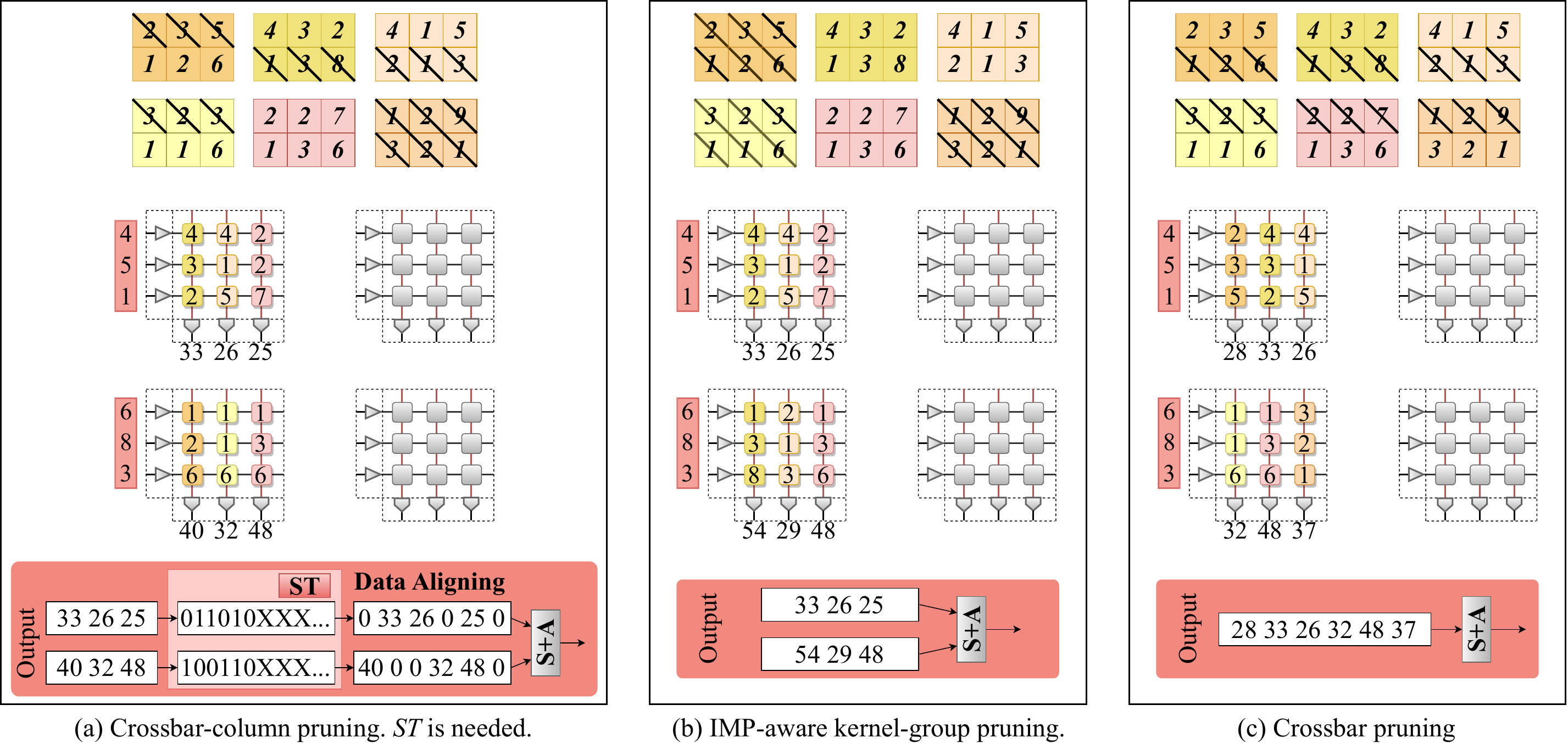} 


\caption{The execution of a convolutional layer with three types of sparsity: crossbar-column, kernel-group, and crossbar. Crossbar-column pruning needs the data alignment hardware unit -- Sparsity Table (\textit{ST}).
} \label{fig:whole-model}

\end{figure}

Pruning methods have been proposed to reduce model complexity and fit complex DNN models into the limited crossbars of the IMP architecture. The current IMP-aware pruning methods are fine-grained and focus on crossbar column/row pruning. For instance, XBA \cite{crossbaraware} prunes the columns of weights within a crossbar and then recombines the weights from different sparse crossbars to reduce the number of used crossbars. Similarly, SPRC \cite{structuredPruning} introduces a multi-group Lasso method that prunes a group of columns of weights in a crossbar. And this method still relies on crossbar-column pruning. PIM-Prune \cite{chu2020pim} proposes to exploit the sparsity in both row and column directions of the weight matrix and designs a new hardware data path to support their pruning method. Nonetheless, these fine-grained methods require hardware masks to align the output/input data of crossbars to function correctly, which introduces significant hardware overhead \cite{structuredPruning, chu2020pim}. To produce the necessary mask, PIM-Prune \cite{chu2020pim} proposes to use the Sparsity Table (\textit{ST}). Take the crossbar-column pruning on the layer of Fig.\ref{fig:mapping} (a) for an example, as shown in Fig.\ref{fig:whole-model} (a), each $1$ in \textit{ST} represents picking a number from the original crossbar output, and $0$ means inserting a $0$ into the final output queue. Therefore, each crossbar requires a corresponding mask memory in the \textit{ST} \cite{chu2020pim}. Meanwhile, the length of a mask in an \textit{ST} equals the maximum pruning rate times the number of weight columns in each crossbar. For example, the \textit{ST} size calculated in Fig.\ref{fig:cimpara} uses a maximum pruning rate of $32$ and each crossbar is configured with $32$ columns of weights, resulting in a \textit{ST} of $1$KB for $8$ crossbars and introducing about $44\%$ more memory and $10\%$ more area into an In-Situ Multiply Accumulate (\textit{IMA}) \cite{shafiee2016isaac}, and the mask may support a higher pruning rate. Also, supporting crossbar-row pruning requires another \textit{ST} to align the input \cite{chu2020pim}, further exacerbating the overhead and making it unaffordable. Moreover, varying pruning rates among different crossbars may result in a significant waste of resources in the \textit{ST} meant for the maximum pruning rate.

Meanwhile, quantization is an approach to reducing the complexity of DNNs and removing FP operations from DNNs for IMP architectures. Traditional quantization methods \cite{jacob2018quantization, choi2019accurate} train a quantized model to determine the appropriate scaling factors that minimize the accuracy loss. PACT \cite{choi2019accurate} optimizes the input clipping parameter during training to identify the appropriate quantization scale for the input, while for weight quantization, they use statistics-aware weight binning to determine the optimal scaling factor based on the statistical characteristics of the weight distribution. However, the scaling factors derived by this method are still in FP format, which necessitates the use of FP processors. IAO \cite{jacob2018quantization}, in addition to training for optimal scaling factors, also proposes a scheme to eliminate the need for FP scaling factors by using fixed-point multiplication. However, this approach still requires fixed-point multipliers, which introduce overhead to the hardware. As shown in Fig.~\ref{fig:cimpara}, each multiplier introduces about $7\%$ power overhead and $9\%$ area overhead into each \textit{Tile} \cite{shafiee2016isaac}, leading to lower energy efficiency and hardware integration density, longer pipeline cycles and worse performance. Consequently, it is crucial to eliminate all FP scaling factors and only incorporate integers in the quantized DNN model.

ReRAM-based IMP accelerators offer notable advantages in terms of area and power efficiency after pruning and quantization of the inferred DNNs. However, ReRAM suffers from various non-idealities in the analog domain, leading to parameter deviations in pre-trained DNN models and further resulting in a significant decrease in inference accuracy. One prevalent non-ideality is SAFs \cite{chen2014rram}, where some device cells are consistently in either a high-resistance state (i.e., SF1) or a low-resistance state (i.e., SF0). It is a kind of error from the fabrication process and is hard to avoid. The previous work \cite{chen2014rram} has reported that SF1 and SF0 can affect about $9.04\%$ and $1.75\%$ of the total device cells, respectively.  Another non-ideality is WV \cite{hsu2015study}, which encompasses cycle-to-cycle variations (CCV) and device-to-device variations (DDV). DDV can be improved with precise manufacturing control and fabrication process advancements, but CCV is an intrinsic property of ReRAM resistive switching behavior and is caused by the stochastic nature of the formation and rupture of a conductive filament, i.e., oxygen vacancies generation and migration process \cite{charan2020accurate}. Previous works have investigated huge amounts of stochastic write variations and found that the variations follow a lognormal distribution \cite{charan2020accurate, long2019design}, indicating that the ReRAM resistance variations follow Eq. (\ref{eq:1}), where $r$ is the ideal resistance value to be programmed, $r'$ is the actual value programmed, and $\theta \sim N(0, \epsilon^2)$, $N$ is the normal distribution with a mean of $0$ and variance of $\epsilon^2$. We also employ this formulation to simulate the WV in this work.
\begin{equation}
\label{eq:1}
    r' \leftarrow e^{\theta}\cdot r, \quad \theta \sim N(0, \epsilon^2)
\end{equation}


Previous works have attempted to mitigate the impact of non-idealities in ReRAM-based IMP accelerators from both software and hardware perspectives. AIIR \cite{charan2020accurate} proposed a joint algorithm-design solution that combines knowledge distillation and random sparse adaptation, but it introduces some hardware overhead. RVComp \cite{he2023rvcomp} aimed to compensate for non-idealities by utilizing extra ReRAM cells to represent the difference between ideal and actual resistance values, also resulting in hardware overhead. Not only do these methods result in less integration density but also more operational stages are introduced into the inference stage, lowering the efficiency. On the other hand, CorrectNet \cite{eldebiky2022correctnet} and DRD \cite{long2019design} utilized non-idealities-aware training methods to reduce the impact of non-ideal ReRAM cells by software algorithms, but they neglect the execution flow of model inference on ReRAM, disregarding the influence of factors such as crossbar size,  ADCs, and others. Also, they do not consider the interaction with IMP-aware pruning. Take WV as an example, it can be inferred from Eq. (\ref{eq:1}) that a weight of zero does not incur any errors. And pruning algorithms can set some weights to zero for compression.  Then these methods for processing WV can be equivalent to pruning, that is, setting many weights to zero. This process reduces the number of weights that can introduce errors, as zero weight does not participate in the computation and does not introduce any errors, thereby decreasing the impact of WV. VACTSF \cite{huang2022rescuing} also tries to remove weights from the model during processing the WV, but a huge accuracy drop ($6.46\%$) is introduced by this method. To improve inference accuracy while maintaining the benefits of pruning, we need to investigate the co-optimization of our model pruning and non-idealities adaptation strategies.

\section{Methodology}
\label{section:methodology}

In this section, we present the details of our methods that optimize DNN models for IMP architectures. The first method is a crossbar-aligned pruning algorithm that combines kernel-group pruning and crossbar pruning to achieve significant sparsity without hardware overhead. The second method is an integer-only quantization approach that eliminates the need for multipliers. The third method is a runtime-aware non-ideality adaptation scheme that considers the IMP inference flow. Lastly, we introduce a DNN learning framework that co-optimizes these schemes in a single training process and extracts the optimal DNN architecture and weights to improve accuracy.

\subsection{Crossbar-Aligned DNN Pruning}
\label{subsection:pruning}
Current IMP-aware pruning methods require extra hardware for data alignment, as shown in Fig.~\ref{fig:whole-model} (a), which reduces integration density and further lowers the number of crossbars on the IMP chip (e.g., the area of the data alignment unit is about $6$ time that of a crossbar array), thus making it difficult to fully benefit from these methods in terms of hardware savings. To solve this problem, we propose a crossbar-alignment pruning algorithm that ensures the pruned DNN model is data-aligned without additional hardware units. Coarse-grained kernel pruning and medium-grained crossbar pruning can maintain data alignment during pruning. However, as reported in prior work \cite{mao2017exploring} that coarser pruning granularity leads to a more decrease in accuracy at the same sparsity level, these fine-grained IMP-aware pruning methods \cite{crossbaraware,enablingReRAM,chu2020pim} result in significant accuracy drops when employed for data-aligned pruning directly. To solve this problem, we introduce a new multi-grained pruning method by combining kernel pruning and crossbar pruning to enlarge the sparsity rate and reduce accuracy loss under data alignment.


\subsubsection{Kernel-group Pruning}

Kernel pruning is a technique that directly removes entire kernels from DNNs. If we consider a DNN model with two connected convolutional layers, denoted as $l_1$ and $l_2$, then the number of kernels in layer $l_1$ is equal to the number of channels in layer $l_2$. We represent the shape of the kernels in these two layers as $C_2 \times K_{1} \times K_{1} \times C_1$ and $C_3 \times K_{2} \times K_{2} \times C_2$, where $C_n$ denotes the number of channels in layer $l_n$, $C_{n+1}$ denotes the number of kernels in layer $l_n$, and $K$ denotes the kernel size.  From Fig.~\ref{fig:mapping} (c), it can be derived that layer $l_1$ requires $\lceil C_2 / XB_w \rceil \times \lceil (K_{1} \times K_{1} \times C_1) / XB_h \rceil$ crossbars, where $XB_w$ and $XB_h$ represent the width and height of a crossbar, respectively. Similarly, layer $l_2$ needs $\lceil C_3 / XB_w \rceil \times \lceil (K_{2} \times K_{2} \times C_2) / XB_h \rceil$ crossbars. By removing kernels, the number of kernels $C_{n+1}$ in layer $l_n$ can be reduced. To maximize the savings of crossbars, we introduce a kernel-group pruning method, as shown in Fig.~\ref{fig:whole-model} (b), which removes a group of kernels from each layer and ensures that the number of remaining kernels in each layer is an integer multiple of $XB_w$. During this pruning process, we first calculate the number of kernels that need to be removed in each layer using the pruning ratio and the importance rank (see Section \ref{subsection:compact}). Then, we round the number of remaining kernels of each layer to integer times of $XB_w$.

Kernel-group pruning has the potential to achieve more crossbar savings. Specifically, let $C_{n+1}^{'}$ represent the remaining kernels in layer $l_n$. Based on the above design, the number of used crossbars in layer $l_1$ and layer $l_2$ are $C_2^{'} / XB_w \times \lceil (K_{1} \times K_{1} \times C_1) / XB_h \rceil$ and $C_3^{'} / XB_w \times \lceil (K_{2} \times K_{2} \times C_2^{'}) / XB_h \rceil$, respectively. If the crossbar is square (i.e., $XB_w = XB_h$) or $(K_{2} \times K_{2} \times XB_w) \mid XB_h$, the by-product of kernel-group pruning in layer $l_1$ can guarantee that the kernels in layer $l_2$ can fully occupy crossbars from the rowing aspect, leading to maximum crossbar savings. In other cases, the by-product of kernel-group pruning in layer $l_1$ can still save some crossbars of layer $l_2$ from the rowing aspect. However, coarse-grained pruning methods, like kernel-group pruning, result in a larger accuracy drop than medium/fine-grained pruning methods, especially under a large pruning ratio \cite{mao2017exploring}. To preserve accuracy, the amount of crossbar savings achieved through kernel-group pruning is limited, indicating the potential for further compression of the pruned model.

%

\subsubsection{Crossbar Pruning}

Crossbar pruning involves removing the crossbar-block of weights from the DNN model to save crossbars, as depicted in Fig.~\ref{fig:whole-model} (c). As a medium-grained unstructured pruning method, crossbar pruning is exclusive to crossbar architecture. By removing a whole crossbar at once, data alignment units are not required in crossbar pruning. Moreover, crossbar pruning can compress both convolutional and fully-connected layers, unlike kernel pruning, which can only compact convolutional layers. Based on the mapping scheme, a mask layer is required after each convolutional or fully-connected layer for crossbar pruning, where each mask value is associated with a crossbar, and zeroizing a mask value means removing all weights in this corresponding crossbar. Mask layers are differentiable to the loss function, hence optimized to reduce loss during the training stage. After training, crossbar-blocks of weights with zerorized mask values are removed from the DNN model (see Section \ref{subsection:compact}), and other mask values are multiplied to corresponding weights before inference to eliminate the mask operation in the inference stage.

Although crossbar pruning can achieve higher sparsity than kernel-group pruning, the mask layer poses challenges to the training stage. These challenges primarily stem from two aspects. First, deeper DNNs are challenging to train, as various layers in the DNN tend to learn at different rates \cite{nielsen2015neural}. The mask layer increases the difficulty of model training to converge. Second, in the zero-recovery training process (see Section \ref{subsection:compact}), more mask values lead to a larger variation of the compressed model architecture during training, intensifying the difficulty of convergence. Therefore, we propose to use kernel-group pruning to compress the DNN first to reduce the number of mask values, reducing the impact of mask layers in the crossbar pruning for higher accuracy.

\subsection{Integer-Only DNN Quantization}
\label{subsection:fp}


Quantization is a technique used to represent the FP inputs (activations) and weights in DNNs using integers with $n$-bits. However, this technique typically involves non-integer scaling factors to adjust the range of inputs and weights, which are necessary to achieve better accuracy \cite{choi2019accurate, jacob2018quantization}. These scaling factors require the use of multipliers, which can introduce significant power and area overhead into the IMP device, as demonstrated in Section \ref{section:relatedwork}. Given the increasing need for high integration density \cite{crossbaraware}, the area of the multiplier, however, can be up to $11$ times that of the crossbar. Therefore, eliminating multipliers from IMP architectures can significantly increase the number of crossbars, thereby enabling the support of more complex DNN models.





To remove FP from DNNs for IMP architectures, we adopt quantization for both inputs and weights. Specifically, since the voltage signals in IMP architectures can represent both positive and negative integers, we adopt a symmetric quantization scheme for inputs. However, as the conductance of the crossbar cells can only be positive, we employ an asymmetric quantization scheme for weights. For instance, in the case of a convolutional layer, the quantization process can be expressed by Eq. (\ref{eq-quan1}), where $X^L$, $W^L$, and $B^L$ denote the inputs, weights, and biases of the layer $L$, respectively, $S$ represents the scaling factor, $Q$ denotes the corresponding quantized value, and $Z$ denotes the zero-point. And all values, except for $Q_X$, are determined during the training process and remain constant during inference. In the inference stage, the input $X^{L+1}$ of the next layer is computed as $W^L \circledast X^L + B^L$ (with  Batch Normalization (BN) layers fused into the convolutional layer \cite{jacob2018quantization} if BN layer exists, same to mask layers), as shown in Eq. (\ref{eq-quan2}). Meanwhile, the scaling factor $S_{X^{L+1}}$ of the next layer is also determined during training, and $X^{L+1}$ can be represented as $X^{L+1} = S_{X^{L+1}}Q_{X^{L+1}}$. However, the crossbar in IMP architectures only supports integer arithmetic operations ($Q_{X} \circledast (Q_{W}-Z_{W})$, $Q_{B}-Z_{B}$), making it necessary to derive the quantized value $Q_{X^{L+1}}$  in Eq. (\ref{eq-quan3})  for performing the next layer's inference in the crossbar.

\vspace{-0.35cm}
\begin{equation}
\label{eq-quan1}
\begin{aligned}
    X^L = S_{X^L}Q_{X^L},\ W^L = S_{W^L}(Q_{W^L} - Z_{W^L}),\ B^L = S_{B^L}(Q_{B^L} - Z_{B^L} )
\end{aligned}
\end{equation}



\begin{equation}
\label{eq-quan2}
\begin{aligned}
    X^{L+1} = W^L \circledast X^L+B^L= S_{X^L}S_{W^L}Q_{X^L} \circledast (Q_{W^L}-Z_{W^L}) +S_{B^L}(Q_{B^L}-Z_{B^L})
\end{aligned}
\end{equation}

\begin{equation}
\label{eq-quan3}
\begin{aligned}
    X^{L+1} = S_{X^{L+1}}Q_{X^{L+1}} => Q_{X^{L+1}} = \frac{S_{X^L}S_{W^L}}{S_{X^{L+1}}}Q_{X^L}  \circledast (Q_{W^L}-Z_{W^L}) + \frac{S_{B^L}}{S_{X^{L+1}}}(Q_{B^L} - Z_{B^L}) \\
\end{aligned}
\end{equation}
To minimize the power and area overhead associated with FP scaling factors in IMP devices, we propose to approximate these factors by powers of $2$, as shown in Eq. (\ref{eq-quan4}). This approach allows for the calculation of scaling factor multiplications using bit-shift units in crossbars. Unlike the widely used representation of FP values as $a\cdot2^b$ ($1 \leq |a| < 2$) in computers under the \textit{IEEE Standard 754} \cite{kahan1996ieee}, approximating $a = 1$ in our method will sacrifice some accuracy. To mitigate the loss of accuracy in our method, we integrate it into the learning process (see Section \ref{subsection:compact}) to enable its co-optimization with the crossbar-aligned pruning and the runtime-aware non-ideality adaptation.

\vspace{-0.3cm}
\begin{equation}
\label{eq-quan4}
\begin{aligned}
   \text{Force\ \ } S_{X^L} &= 2^{a1}, S_W = 2^{a2}, S_B = 2^{a3}, S_{X^{L+1}} = 2^{b1} \\
    Q_{X^{L+1}} = 2&^{a1+a2-b1}Q_{X^L} \circledast (Q_{W^L}-Z_{W^L}) + 2^{a3-b1}(Q_{B^L}-Z_{B^L})
\end{aligned}
\end{equation}




\subsection{Runtime-aware Non-Ideality Adaptation}
\label{subsection:variation}


The aim of training a DNN is to find the global minimum of the loss function. However, due to the flexibility of the DNN parameters, there exists a region where the loss function remains small despite not being the global minimum \cite{long2019design}. Within such a region, the DNN's parameters can vary without significantly impacting the accuracy. To create a DNN that is robust to ReRAM non-ideality, we inject the non-ideality into the training process. This ensures that the trained DNN is resilient to non-ideality, improving accuracy in non-ideal ReRAM-based IMP accelerators. 

We first analyze the actual inference process of DNN models on the ReRAM-based crossbar, and the error introduced in this process. In this analysis, each ReRAM cell can store a $2$-bit weight and each weight of the DNN model is quantized into $8$ bits and then mapped to $4$ ReRAM cells. The input of the DNN model is also quantized into $8$ bits and injected into the crossbar bit by bit (each bit is represented by $v$, $v \in \{0,1\}$). The height of each crossbar is $H$, and the comparator in ADC can round the output of each column in the crossbar into an integer-format digital value. For a dot product operation of vector-vector multiplication (VVM) mapped onto the crossbar (like \{4 5 1 6 8 3\} $\cdot$ \{2 3 5 1 2 6\}, as shown in Fig.~\ref{fig:mapping} (c)), the weight vector is mapped onto the same column of several crossbars, and the input vector is converted to voltage and injected into the wordline of the crossbar. Assume the dimensions of these two vectors are both $D$, and $H$ is the height of each crossbar, then the number of used crossbars is $\lceil D/H \rceil$. For simplicity, we assume that  $\lceil D/H \rceil = D/H$, which means that the vector can fully occupy the used crossbar. Otherwise, if $\lceil D/H \rceil > D/H$, then some  ReRAM cells are not used and programmed to $0$, and the error should be smaller. Take WV as an example, we use $c$ to represent the ideal value to be programmed onto the ReRAM cells ($c \in \{0,1,2,3\}$, $c=1/r$, $r$ is the resistance),  and $c'$ is the actual value programmed (according to Eq. (\ref{eq:1}), $c'=e^{-\theta}\cdot c$; considering $\theta$ is in a normal distribution, we also use $c'=e^{\theta}\cdot c$.). Eq.~\ref{v-1} shows the ideal calculation process for a VVM operation. When taking WV into consideration, the calculation process becomes Eq.~\ref{v-2}. The error introduced by this process is given by Eq.~\ref{v-3}.


\vspace{-0.2cm}
\begin{equation}
\label{v-1}
    Results = \sum_{d=1}^{D/H} (\sum_{h=1}^{H} v_{dh}\cdot c_{dh}), 
\quad \quad v_{dh} \in \{0,1\}, \quad  c_{dh} \in \{0,1,2,3\}
\end{equation}

\begin{equation}
\label{v-2}
    Results' = \sum_{d=1}^{D/H} Round(\sum_{h=1}^{H} v_{dh}\cdot c'_{dh}) = \sum_{d=1}^{D/H} Round(\sum_{h=1}^{H} v_{dh}\cdot e^{\theta}\cdot c_{dh})
\end{equation}

\begin{equation}
\begin{aligned}
    \label{v-3}
    Error = |Results' - Results| &= | \sum_{d=1}^{D/H} Round(\sum_{h=1}^{H} v_{dh}\cdot e^{\theta}\cdot c_{dh}) - \sum_{d=1}^{D/H} (\sum_{h=1}^{H} v_{dh}\cdot c_{dh})| \\
    &= | \sum_{d=1}^{D/H} Round(\sum_{h=1}^{H} v_{dh}\cdot (e^{\theta} - 1) \cdot c_{dh})|
\end{aligned}
\end{equation}

To quantify the magnitude of the error introduced by the WV, we calculate the mathematical expectation of the error. As the input value ($v$), the weight value ($c$), and the WV of ReRAM cells ($\theta$) \cite{charan2020accurate} are independently distributed, the mathematical expectation of the error is shown in Eq.~\ref{v-4}. In the quantized DNN models, we assume that each bit of the input ($v$) is $0$ or $1$ with equal probability (inject one bit of input into the crossbar each time), and the weight ($c$) is in \{0,1,2,3\} with the same probability. Then $E(v_{dh}) = E(v) = 0.5$ and $ E(c_{dh}) = E(c) = 1.5$. And the expectation of the lognormal distribution is $E(e^\theta) = e^{\epsilon ^{2}/2}$. Thus, the expectation of the error is simplified to Eq.~\ref{v-5}. 

\begin{equation}
    \label{v-4}
    E(Error) = |\sum_{d=1}^{D/H} Round(\sum_{h=1}^{H} E(v_{dh}) \cdot (E(e^{\theta}) - 1) \cdot E(c_{dh}))|
\end{equation}

\begin{equation}
\begin{aligned}
    \label{v-5}
    E(Error) &= \sum_{d=1}^{D/H} |Round(\sum_{h=1}^{H} 0.5 \cdot (e^{\epsilon ^{2}/2} - 1) \cdot 1.5)| \\
    & = \frac{D}{H} \cdot Round({H} \cdot 0.75 \cdot (e^{\epsilon ^{2}/2} - 1)), \quad E(e^\theta) = e^{\epsilon ^{2}/2} \geq 1
    \end{aligned}
\end{equation}

\begin{figure}[!t]



    \includegraphics[trim={0pt 0pt 0pt 0pt},clip,width=0.99\linewidth]{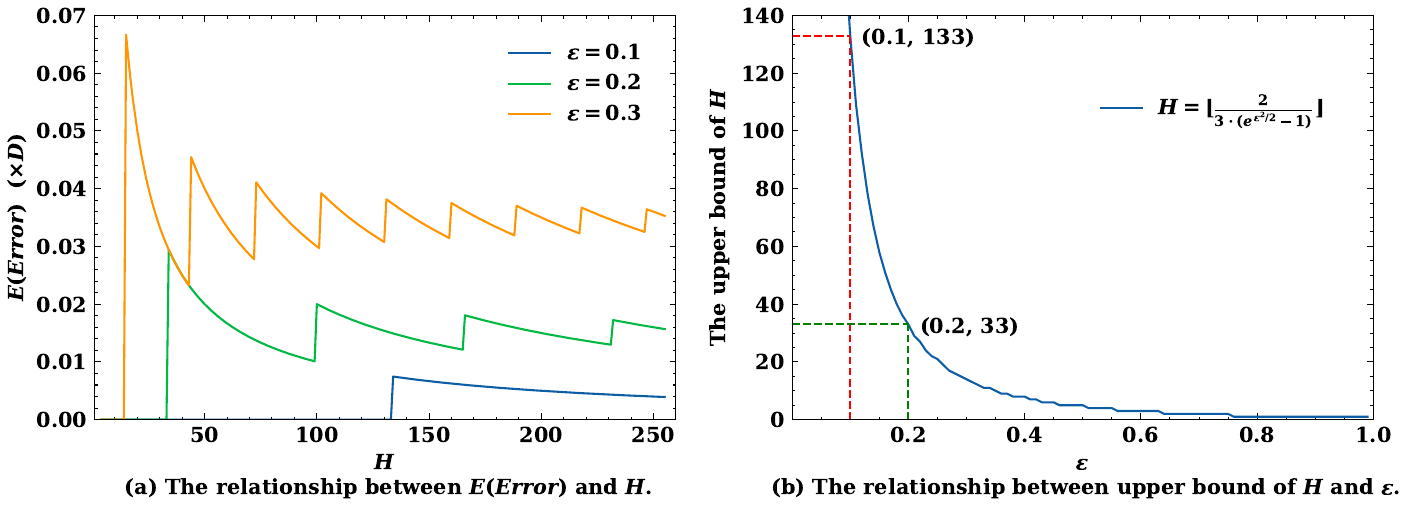} 


\caption{Relationship between error expectation ($E(Error)$), the crossbar height ($H$), and the WV degree ($\epsilon$).
} \label{fig:error_h_relation}
\end{figure}

Therefore, from Eq.\ref{v-5}, we can deduce that $E(Error)$ is related to the height ($H$) and WV ($\epsilon$) of the crossbar. Fig.\ref{fig:error_h_relation} (a) depicts the relationship between $E(Error)$ and $H$ under different values of $\epsilon$ according to Eq.\ref{v-5}.  From this figure, we can more intuitively find that when ${H} \cdot 0.75 \cdot (e^{\epsilon ^{2}/2} - 1) < 0.5$, which is equivalent to ${H} < \frac{2}{3\cdot(e^{\epsilon ^{2}/2} - 1)}$, then $E(Error) = 0$. This means that when $H$ is small, the error accumulated in each crossbar is insignificant, and it can be eliminated through the round function in the ADC. On the other hand, when $n \leq {H} \cdot 0.75 \cdot (e^{\epsilon ^{2}/2} - 1) < n+0.5$, where $n=0,1,2,...$, $E(Error) = \frac{nD}{H}$. This function decreases as $H$ increases. Therefore, $H$ should be increased to the upper bound, which is ${H} = \lfloor \frac{2+4n}{3\cdot(e^{\epsilon ^{2}/2} - 1)}\rfloor, \epsilon \neq 0$.  Also, the function graph of ${H} = \lfloor \frac{2}{3\cdot(e^{\epsilon ^{2}/2} - 1)}\rfloor, \epsilon \neq 0$ is shown in Fig.~\ref{fig:error_h_relation} (b), and the region below the blue line indicates the valid values of $H$ that guarantee $E(Error) = 0$. For example, if $\epsilon = 0.1$, then $H\leq133$. If $\epsilon = 0.2$, then $H\leq33$. Until now, we establish a relationship among the height of each crossbar, the WV of ReRAM cells, and the mathematical expectation of the error introduced by WV. This demonstrates the necessity of perceiving the runtime inference process during the non-ideality adaptation, which enables the error to diminish only through the comparator in the ADC, without incurring any hardware overhead. In addition, this finding serves as a useful reference not only for supporting the algorithm in this paper but also as a guideline for the further design of ReRAM-based IMP devices. Similarly, the above method can also be applied to analyze the error introduced by SAF, etc.

\begin{figure}[!t]



    \includegraphics[trim={0pt 0pt 0pt 0pt},clip,width=0.99\linewidth]{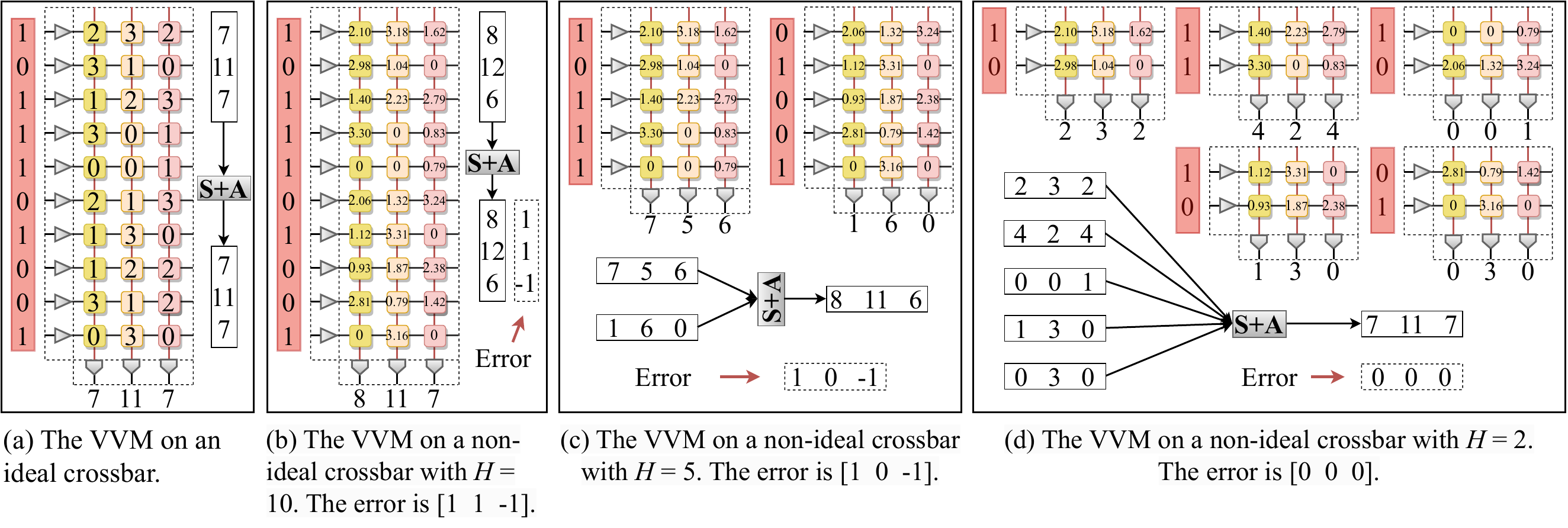} 


\caption{Diagrammatic representation of the effect of crossbar height on the error introduced by WV.
} \label{fig:error_h}
\end{figure}

\begin{figure}[!t]



    \includegraphics[trim={25pt 0pt 0pt 0pt},clip,width=0.99\linewidth]{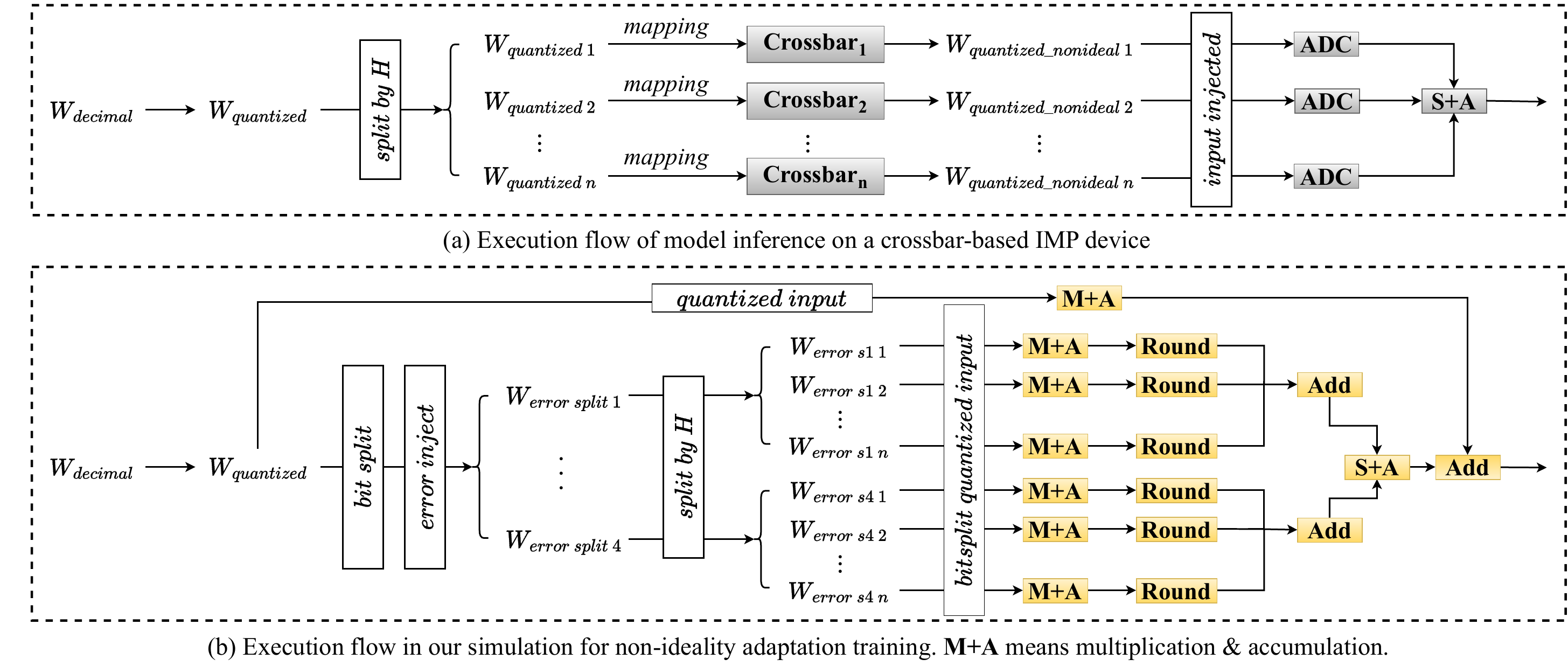} 


\caption{Execution flow of DNN on ReRAM IMP and its corresponding simulated flow.
} \label{fig:flow}
\end{figure}

The height of the crossbar $H$ plays a significant role in the error introduced by WV in the IMP device. When $H$ is small, the round function by the comparator in ADC can automatically correct the error, as illustrated in Fig.~\ref{fig:error_h}. This characteristic can be leveraged in the non-ideality adaptation training process to achieve high accuracy. Each ReRAM cell we used in the IMP device can store only $2$-bit parameters. For mapping an $8$-bit weight parameter, $4$ adjacent cells are required. Since the height of the crossbar in the IMP device has a great influence on the error, directly multiplying DNN parameters multiplied by $e^\theta$ as used in other methods \cite{long2019design} is not applicable for simulating the error of WV.  Fig.~\ref{fig:flow} (a) illustrates the inference process of the DNN model on a ReRAM-based IMP device. First, the $32$-bit decimal parameters ($W_{decimal}$) are quantized into $8$-bit integers ($W_{quantized}$), and then these quantized parameters are mapped and programmed into the ReRAM crossbar, according to Fig.~\ref{fig:mapping} (c). If the height of the parameter matrix is greater than the height of the crossbar, the parameter matrix is split into $n$ smaller matrices, with each matrix mapped to one crossbar. During this mapping process, the non-ideality of the ReRAM cells leads to the programmed values ($W_{quantized\_{nonideal}}$) differing from the required values. The multiplication of the non-ideal weights and inputs is then calculated on the crossbar, and the output of each crossbar is converted into integer digital values by the ADC. Finally, the digital output of each crossbar is shifted and added by the \textit{S+A} unit to obtain the final result.

To exploit the crossbar characteristics described above for higher accuracy under device non-ideality, it is crucial to accurately simulate the execution flow of the DNN on the ReRAM crossbar. Fig.~\ref{fig:flow} (b) depicts the simulation process for DNN runtime inference on the crossbar (named runtime simulator), which also begins by quantizing the decimal weight matrix ($W_{decimal}$) into $8$-bits integers ($W_{quantized}$). The multiplication and accumulation (\textit{M+A}) are then computed between the quantized weight and the quantized input, to obtain the ideal output without considering device non-ideality. To account for the non-ideality introduced, each quantized weight of $8$ bits is divided into $4$ parts, each corresponding to the ideal mapped value on a ReRAM cell of the crossbar. We then multiply each part of the split weight by $e^\theta-1$ to obtain the error of weight introduced by non-ideality ($W_{error}$). Next, we split the error of weights according to the height of the crossbar ($H$) to simulate the calculation process on each crossbar. The multiplication and accumulation between the error of weights and the bit-split quantized input are then computed, and the result is rounded to integer format. By combining all outputs from all splits, we obtain the final error, which is added to the ideal result to obtain the equivalent result of running on the crossbar with device non-ideality.

The purpose of this process is to calculate all the errors introduced by non-ideality in parallel while accurately simulating the crossbar operation, which greatly improves efficiency. Additionally, the non-ideality and the ideal result are calculated separately, so that the non-ideality does not affect the training process. This is because the non-ideality is random, and introducing random numbers into the training process will hinder convergence.

\subsection{Comprehensive DNN Learning Framework}
\label{subsection:compact}

The pruning and quantization techniques employed in this work involve using ``large-accuracy-loss" methods to eliminate all extra hardware for completely low-power IMP acceleration. To further minimize the accuracy loss incurred by these methods,  besides the multi-grained pruning scheme, we propose a model learning framework that optimizes the weights and architecture during the training process. Additionally, we incorporate the non-ideality of the ReRAM cells into the training process to enhance the adaptability of the model to non-idealities. During the forward propagation stage, our runtime simulator is used to imitate the actual DNN inference on the device under non-ideality, bringing the non-ideality into the loss function to adapt. Thus, this framework combines pruning, quantization, and non-ideality adaptation together into one training procedure, named comprehensive DNN learning framework, allowing these operations to co-optimize for higher accuracy. Algorithm \ref{code:al1} describes the proposed framework, which is divided into two phases: Kernel-group pruning (\textit{Line 18}) and Crossbar pruning (\textit{Line 20-21}). The quantization and non-ideality adaptation schemes are inserted into each phase for co-optimization.

In crossbar-aligned pruning, a crucial step is to determine which parts of the model can be removed without sacrificing much accuracy. In this regard, an explicit learning medium $\delta$ is required to distinguish the importance of the weight. For kernel-group pruning, we remove entire kernels at once, and we can leverage the trainable parameter $\gamma$ of the BN layer to provide relevant instructions without any overhead, as $\gamma$ can indicate the importance of kernels \cite{zerobn}. And for crossbar pruning, we design a mask layer to distinguish the importance of crossbar-size weights. Regarding the integer-only quantization and non-ideality adaptation techniques,  they do not directly remove or modify weights from the DNN model and they need to quantize or introduce non-ideality into every weight. Therefore, there is no need to distinguish weights in these two techniques.

\begin{algorithm}[t]

\small
    \caption{Comprehensive DNN Learning Framework for IMP}
    \label{code:al1}
    \KwIn{model, training data and settings, zero start epoch: $s$, prune\_ratio: $p$, xb\_size: $XB_s$, quan\_bits: $q$, non\_ideality: $\theta$ }
    \KwOut{the well-trained, compact, and reliable model}
      \DontPrintSemicolon
    \SetKwFunction{FMain}{Comprehensive\_Learning}
    \SetKwProg{Fn}{Function}{:}{}
    \Fn{\FMain{$model$, $\delta$}}{
        
        
        
        \For{$t\gets 1$ \KwTo $epoch_{half}$}{
        
            
            \uIf{$t \geq s$ and ($t \% 2 == 0$ or $t == epoch_{half}$)}{
                $scaling\_factor \gets quantize(model, q)$; \textbf{\textit{\# Prepare quantization}}\;
                Force all scaling factors to $2^n$; \textbf{\textit{\#Integer-Only quantization}}\;
                Calculate $ Weight\_quan$ according to the new $scaling\_factor$; \textbf{\textit{\#Integer-Only quantization}}\;             
                $forward\_with\_simulator(scaling\_factor, weight\_quan, \theta, XB_s)$; 
                \textbf{\textit{\#Training according to  Fig.~\ref{fig:flow}};}\;
                            $update\_with\_sparsity(model, \delta)$; \textbf{\textit{\#Sparse training for pruning}}\;
                $Imp_{r} \gets$ Sort kernels/Crossbars by $|\delta|$; \textbf{\textit{\# Prepare pruning}}\;
                Find $\delta$ threshold $\delta_{th}^l$ of each layer by $Imp_{r}$, $p$, $XB_s$;\textbf{\textit{\# Crossbar-aligned pruning}}\;
                Zerorize $\delta_i^l$ if $\delta_i^l$ $<$ $\delta_{th}^l$;\textbf{\textit{ \#Temporarily removing weights for pruning}}\;
        
            }
            \uElse{
                        $forward(model)$; \textbf{\textit{\#Training}}\;
            $update\_with\_sparsity(model, \delta)$; \textbf{\textit{\#Sparse training for pruning}}\;
            }
        
        }
        Remove all weights from the model with zero $\delta_i^l$;\;

  }
 Initialize the model and its parameters randomly;\;

\textbf{\textit{\#Phase 1 -- Kernel-group pruning}}\;
Comprehensive\_Learning(model, $\gamma$);\;
\textbf{\textit{\#Phase 2 -- Crossbar pruning}}\;
model\_mk $\leftarrow $ Mask(model);\;
Comprehensive\_Learning(model\_mk, mask);\;

\end{algorithm}

The following is a description of the core function \textit{Comprehensive\_Learning} (\textit{Line 1-15}) in detail. Initially, the model undergoes training for several epochs, named initial epochs ($t<s$ in \textit{Line 3}, \textit{Line 13-14}), to allow all the learning medium (importance factors) $\delta$ to have informative values that reflect kernel/crossbar importance, rather than being initialized randomly (\textit{Line 16, Line 20}). This also enables the model weights to be trained without the disturbance of pruning, quantization, and weight non-ideality, reaching a region where the loss is relatively small more quickly. Furthermore, we use sparse training \cite{zerobn} (\textit{Line 8, Line 14}) throughout the training process, forcing the importance factors $\delta$ to approach zero for safely removing unimportant weights in the pruning process.

The function next enters the zero-recovery training epochs (\textit{Line 2-14}). In the zero epoch (\textit{Line 4-11}), the first step is integer-only quantization (\textit{Line 4-6}), where the scaling factor is computed based on the quantization bits ($q$, we use $8$) (\textit{Line 4}). These scaling factors are then rounded to the nearest power of $2$ for easy computation on the \textit{S+A} units in the IMP device (\textit{Line 5}). Finally, the quantized weights ($Weight\_{quan}$) are computed using these scaling factors to complete the quantization process (\textit{Line 6}). It should be noted that the model is quantized for inference during the forward propagation stage only, with no changes to the weights, and the backpropagation remains the same as traditional training. The second step is the non-ideality adaptation (\textit{Line 7}), where the forward propagation stage is performed with our runtime simulator to incorporate the non-ideality in the training process for adaptation. Finally, crossbar-aligned pruning is performed, where the importance factors $\delta$ are sorted according to their absolute values to obtain the global importance rank of kernels/crossbars in the entire model (\textit{Line 9}). During the entire training process, these importance factors are jointly optimized with the network weights, and the network can automatically identify the importance of each kernel/crossbar. Next, the threshold of importance factors is calculated for each layer (\textit{Line 10}) to ensure that the compact model fully occupies each used crossbar. Finally, the unimportant parts are temporarily shielded for the current zero epoch by setting $\delta_i^l$ to zero for those parts with importance factors less than the threshold (\textit{Line 11}).


After each zero epoch, we employ a recovery epoch (\textit{Line 3, Line 13-14}) to improve the accuracy of the model. Similar to knowledge distillation, in the zero epoch, the training process aims to use a pruned, quantized, and non-ideality injected model to cover all knowledge from the original model. There are two scenarios in this process: 1) if the loss is small, then the network retains these weights in the whole training process; 2) if the loss is huge, the optimizer significantly updates the weights. In the second scenario, the recovery epoch enables these weights to reach another region where the loss is small without being affected by pruning, quantization, and adaptation. Therefore, the recovery epoch enables the learning process to obtain better weights for high accuracy.

In addition, the recovery epoch also provides an opportunity for the previously zerorized importance factors (i.e., $\delta_i^l$) to recover and potentially play a crucial role in subsequent training and inference processes, allowing our pruning method to find a more efficient model architecture with higher accuracy. Eq. (\ref{eq-zerobn-1}) provides the common calculation formula for a convolutional layer with $\delta$ and an activation function ($\sigma$), where $L$ represents the layer index, $X^L$ represents the input of layer $L$, $W^L$ represents the weight between layer $L+1$ and $L$, and $\sigma^L$ represents the activation function at layer $L$. The symbol $\circledast$ denotes the convolution operation.  During the recovery epoch, we mainly focus on updating the importance factors $\delta^{L}$ that are set to zero during the previous zero epoch. The gradients of $\delta^{L}$ to the final loss function can be calculated using Eq. (\ref{eq-zerobn-2}). Given that some values in $\delta^L$ have been zerorized and the commonly used activation function ($\sigma$) is  \textit{ReLU}, the output of $\sigma^L$ is also zero under \textit{ReLU}. Although \textit{ReLU} is not differentiable at zero, it is commonly assumed that its derivative is also zero. Therefore, we can easily obtain Eq. (\ref{eq-zerobn-4}).

\vspace{-0.3cm}
\begin{equation}
\label{eq-zerobn-1}
 X^{L+1} = \sigma^L(\delta^{L} \cdot (W^L \circledast X^{L} + B^L))
 \end{equation}
\begin{equation}
\label{eq-zerobn-2}
   \frac{\partial loss}{\partial \delta^L } = \frac{\partial loss}{\partial X^{L+1}}\cdot {\sigma^L}^{'} \cdot (W^L \circledast X^{L} + B^L)
 \end{equation}
\begin{equation}
\label{eq-zerobn-4}
  \delta^L = 0 \rightarrow {\sigma^L}^{'} = 0 \rightarrow \frac{\partial loss}{\partial \delta^L } = 0
\end{equation}


Unfortunately, it stacks at a deadlock. To break the deadlock caused by the zerorized $\delta^{L}$, small values can be added to $\delta^{L}$. However, many training algorithms employ momentum to accelerate convergence, which accumulates gradients from previous steps to determine the direction to go. In particular, weight updates with momentum are shown in Eq. (\ref{eq-zerobn-5})-(\ref{eq-zerobn-6}), where $lr$ is the learning rate, $z_t$ is the updated value from the previous step $t$, and $m$ is the accumulation coefficient. This allows the weights $W_t$ to be updated to $W_{t+1}$ by combining the gradients of the current and past steps.

\vspace{-0.3cm}
\begin{equation}
\label{eq-zerobn-5}
    z_{t+1} = m\cdot z_t +  \frac{\partial loss}{\partial W_t}
\end{equation}
\begin{equation}
\label{eq-zerobn-6}
    W_{t+1} = W_t - lr \cdot z_{t+1}	
\end{equation}

As a result, even though the current gradients of $\delta^{L}$ are zero, the proposed framework can update them to recover from zero by following the last updating directions, without requiring additional steps. This process is fully automated and efficient.  Previously zerorized kernels/crossbars may become significant, and they are not zerorized in the future zero epoch, thereby altering the architecture of the final compact model. Furthermore, the training process aims to minimize loss and improve accuracy, which enables the compact architecture to become better and better. In contrast to other IMP-aware pruning methods \cite{crossbaraware,enablingReRAM, chu2020pim}, which solely propose pruning schemes without an optimized learning process, our proposed framework obviously improves accuracy.

Since our pruning, quantization, and adaptation methods are all tied to the loss function, they can be jointly optimized to improve accuracy during the training process. Finally, we eliminate all weights corresponding to zerorized $\delta_i^l$ (\textit{Line 15}), resulting in a highly efficient, integer-only, and reliable model that is well-suited to the IMP architecture while still achieving high accuracy.

\section{Experimental Evaluation}
\label{section:experiment}
In this section, we first present the details of the hardware platforms, benchmark models, and datasets employed in this paper. Then, we evaluate the performance of our integer-only quantization method. Subsequently, we present a comparative analysis of our crossbar-aligned pruning and integer-only quantization approach against state-of-the-art IMP-aware pruning methods in terms of sparsity rate, accuracy, area, and power. Finally, we evaluate the effectiveness of our comprehensive learning framework in achieving efficient, compact, and reliable DNNs for IMP devices.



%
\begin{table}[!t]
\centering
\small



\caption{Accuracy comparison of different quantization methods.}
\label{tab:quan}
\begin{tabular}{cccccc}
\hline
Dataset                   & Network                   & \begin{tabular}[c]{@{}c@{}}Baseline Acc. (\%)  \end{tabular} & Method & \begin{tabular}[c]{@{}c@{}}Quantized Acc. (\%)  \end{tabular} & \begin{tabular}[c]{@{}c@{}}Acc. Drop (\%)  \end{tabular} \\ \hline
\multirow{2}{*}{Mnist} & \multirow{2}{*}{LeNet-5} & \multirow{2}{*}{99.21}                             & IAO \cite{jacob2018quantization}   & \textbf{99.10 }                                                     & \textbf{0.11}                                                \\
                          &                           &                                                             & \ours-Quan   & 99.09                                                       & {0.12} \\ \hline\hline
\multirow{4}{*}{Cifar-10} & \multirow{2}{*}{VGG-16}    & \multirow{2}{*}{93.28}                                      & IAO \cite{jacob2018quantization}   & 93.14                                                       & 0.14                                                \\
                          &                           &                                                             & \ours-Quan   & \textbf{93.39}                                                       & \textbf{-0.11}                                               \\ \cline{2-6} 
                          & \multirow{2}{*}{Resnet-56} & \multirow{2}{*}{93.34}                                      &  IAO \cite{jacob2018quantization}   &     \textbf{93.44}                                                       &          \textbf{-0.10}                                        \\
                          &                          &                                                             & \ours-Quan   & 93.37                                                       & -0.03                                               \\ \hline\hline
\multirow{2}{*}{ImageNet} & \multirow{2}{*}{Resnet-18} & \multirow{2}{*}{69.79}                                      & IAO \cite{jacob2018quantization}   & \textbf{69.54}                                                       & \textbf{0.25}                                                \\
                          &                           &                                                             & \ours-Quan   & 69.48                                                       & 0.31                                                \\ \hline
\end{tabular}
\end{table}
\begin{table}[!t]
\centering
\small



\caption{Accuracy comparison of final compact models without non-ideality.}
\label{tab:pq}
   \setlength{\tabcolsep}{4pt} 
\begin{tabular}{ccccccc}
\hline
Dateset                   & Network                    & Method & Baseline Acc.  (\%)   & Sparisity Rate (\%)   & Final Acc. (\%)     & Acc. Drop  (\%)       \\ \hline
\multirow{2}{*}{Mnist}                    & \multirow{2}{*}{LeNet-5  }     & LSRR \cite{lin2019learning}   & 99.23         & {92.00} & \textbf{99.15} & \textbf{0.08} \\      &       
& \ours-PQ   & 99.21         & \textbf{94.89} & {98.90} & {0.31}  \\ 

\hline \hline
\multirow{5}{*}{Cifar-10} & \multirow{2}{*}{VGG-16}    & CSAO \cite{enablingReRAM}   & 93.30         & 34.90          & 93.10          & 0.20           \\
                          &                            & \ours-PQ   & 93.28         & \textbf{88.25} & \textbf{93.71} & \textbf{-0.43} \\ \cline{2-7} 
                          & \multirow{3}{*}{ResNet-56} & CSAO \cite{enablingReRAM}  & 92.90         & 23.50          & 92.50          & 0.40           \\
                          &                            & SPRC  \cite{structuredPruning}  & -             & 33.40          & 92.80          & -              \\
                          &                            & \ours-PQ   & 93.34         & \textbf{51.36} & \textbf{93.18} & \textbf{0.16}  \\ \hline \hline
\multirow{4}{*}{ImageNet} & \multirow{4}{*}{ResNet-18} & XBA \cite{crossbaraware}    & 69.31         & 24.89          & 66.07          & 3.24           \\
                          &                            & SPRC \cite{structuredPruning}  & 69.76         & 26.41          & 67.82          & 1.94           \\
                          &                            & PIM-P  \cite{chu2020pim}  & 69.76         & 32.41          & 68.67          & 1.09           \\
                          &                            & \ours-PQ   & 69.79         & \textbf{50.50} & \textbf{68.80} & \textbf{0.99}  \\ \hline
\end{tabular}
\end{table}

\subsection{Experiment Setup}

The IMP accelerator used in this paper is based on the ISAAC architecture \cite{shafiee2016isaac}, which is a widely-adopted platform. The crossbar size is set to $128 \times 128$, and each memristor cell in this architecture can store $2$ bits. We quantize the model to $8$ bits, map each weight to $4$ memristor cells and use the \textit{S+A} unit in the device to combine the results from different cells. For components that ISAAC did not provide, such as \textit{ST}, we use CACTI \cite{muralimanohar2009cacti} at $32$nm to model their power and area. The design configurations are simulated using the modified NeuroSim \cite{chen2018neurosim} simulator and under the $32$nm CMOS library. The proposed training framework is implemented using the PyTorch framework \cite{pytorch}. We evaluate the proposed method on representative DNNs: LeNet \cite{lenet}, VGG \cite{vgg}, and ResNet \cite{resnet} on three datasets: Mnist \cite{deng2012mnist}, Cifar-10 \cite{cifar10}, and ImageNet \cite{ILSVRC}. The total training epochs we used are $200$, $160$, and $90$ for Mnist, Cifar-10, and ImageNet, respectively. As introduced in  Section \ref{section:relatedwork}, we set $9.04\%$ of cells to $0$ (high-resistance state) and $1.75\%$ to $3$ (low-resistance state) to simulate the SAF errors. Meanwhile, for the WV error, we employ two commonly used $\epsilon=0.1$ or $\epsilon=0.5$ in other works \cite{charan2020accurate, long2019design, eldebiky2022correctnet} for comparison.  It is noteworthy that, as shown in Fig.~\ref{fig:error_h_relation} (b) when $\epsilon=0.1$ and $H=128$, the error introduced by WV can be significantly mitigated by the round function in ADC.






%
%
%

\subsection{Evaluation of Model Compact w/o Non-ideality}
\subsubsection{Accuracy evaluation of quantized models} The performance evaluation of the proposed integer-only quantization approach (\ours-Quan) is demonstrated in Table \ref{tab:quan}, where we can find that our approach does not lead to a significant decrease in accuracy compared to the baseline (i.e., full precision), and even outperforms full precision in some models. Moreover, under the same training settings, the accuracy of our technique is similar to that of the IAO quantization method \cite{jacob2018quantization}, which does not enforce scaling factors to be a power of $2$. This highlights the effectiveness of our quantization scheme, which avoids FP multiplication operations while maintaining accuracy.

\begin{figure}[!t]



    \includegraphics[trim={0pt 0pt 0pt 0pt},clip,width=0.90\linewidth]{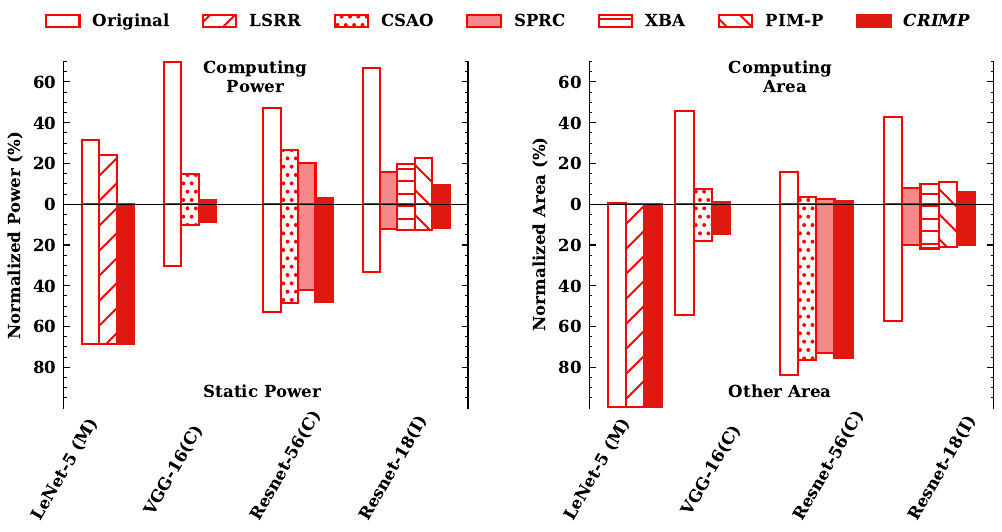} 

\caption{Evaluation of the power and area consumption of different models from different methods.} \label{fig:powerarea}
\end{figure}

\subsubsection{Accuracy evaluation of final compact models}
Table \ref{tab:pq} presents a comparison of our proposed crossbar-aligned pruning method (with integer-only quantization, represented by \ours-PQ) to state-of-the-art IMP-aware pruning methods, including LSRR \cite{lin2019learning}, CSAO \cite{enablingReRAM}, SPRC \cite{structuredPruning}, XBA \cite{crossbaraware} and PIM-P \cite{chu2020pim}. And in these methods, the effect of non-ideality is ignored. It should be noted that other methods still use FP scaling factors in their quantization schemes. The highest sparsity rate, highest accuracy, and lowest accuracy reductions are marked in \textbf{boldface}. From this table, we can find that for small datasets such as Mnist and Cifar-10, all methods can compress the model to a high sparsity rate with only a little accuracy drop. Our method tries to remove more redundant weights from LeNet-5 on the Mnist dataset, which causes a larger accuracy drop than LSRR \cite{lin2019learning}. However, our method can achieve a higher sparsity rate on the VGG-16 model with even accuracy improvement, indicating that VGG-16 overfits the Cifar-10 dataset. For Resnet-56, we can still minimize the accuracy drop \footnote{SPRC does not report the baseline accuracy of Resnet-56 on Cifar-10.}. For a large dataset like ImageNet, the sparsity rate achieved by all methods is relatively lower, but our method achieves a larger sparsity rate and higher accuracy. Overall, our proposed framework achieves comparable or better performance compared to state-of-the-art IMP-aware pruning methods, demonstrating its effectiveness in achieving compact and accurate DNNs for IMP devices without any hardware overhead.

\begin{figure}[!t]



    \includegraphics[trim={0pt 0pt 0pt 0pt},clip,width=0.95\linewidth]{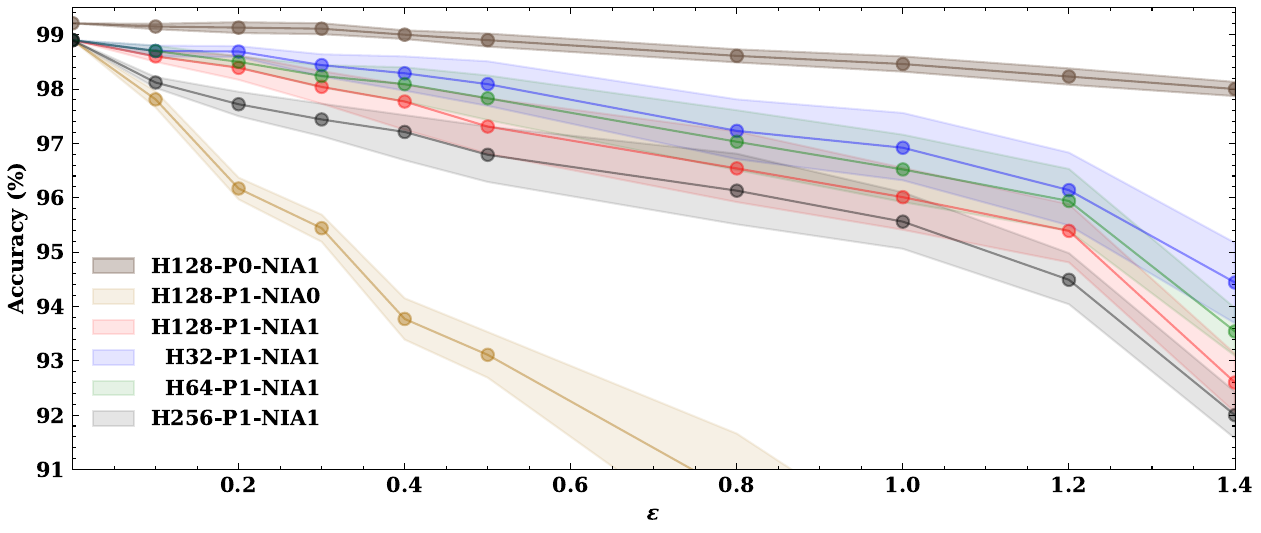} 


\caption{Relationship between accuracy and $\epsilon$ under different crossbar heights (H), with or without pruning (P1 or P0), and with or without non-ideality adaptation (NIA1 or NIA0).
} \label{fig:acc-h-sigma}
\end{figure}

\subsubsection{Power \& area evaluation of final compact models} 
We employ the modified NeoroSim simulator to measure the power and area consumption of models of various methods. It should be noted that crossbar-column and crossbar-row pruning methods require the Sparsity Table unit for data alignment, and methods using FP scaling factors require FP processors.  LSRR \cite{lin2019learning} and SPRC \cite{structuredPruning} use $4$-bit quantization \footnote{We use $4$-bits quantization to model their power and area.}. We normalize the power and area consumption to the total consumption of the original model. And we calculate the computing power and static power separately, as well as the computing area (crossbars, multipliers, and STs) and other areas (interconnect and memory) separately. From Fig.~\ref{fig:powerarea}, we can see that our method achieves the highest power and area efficiency improvement, especially for the computing parts. Compared to the original model, our method achieves an average total power reduction of $4.44\times$ and computing power reduction of $122.38\times$, as well as a total area reduction of $3.16\times$ and computing area reduction of $19.51\times$.

\begin{table}[!t]
\centering
\small



\caption{Evaluation of our final compact \& reliable models for IMPp; A. means \textit{Area}.}
\label{tab:all}
   \setlength{\tabcolsep}{0.3pt} 
\begin{tabular}{cccccccccccc}
\hline
Dateset                    & Network                    & Method & \begin{tabular}[c]{@{}c@{}}Quan.\\ Level\end{tabular} & SAF   & \begin{tabular}[c]{@{}c@{}}Varia.\\ ($\epsilon$)\end{tabular} & \begin{tabular}[c]{@{}c@{}}Sparisity\\ Rate (\%)\end{tabular} & \begin{tabular}[c]{@{}c@{}}Hardware\\ Overhead\end{tabular} & \begin{tabular}[c]{@{}c@{}}Runtime\\ Overhead\end{tabular} & \begin{tabular}[c]{@{}c@{}}Baseline\\ Acc. (\%)\end{tabular} & \begin{tabular}[c]{@{}c@{}}ReRAM\\ Acc. (\%)\end{tabular} & \begin{tabular}[c]{@{}c@{}}Acc. ↓\\ (\%)\end{tabular} \\ \hline
\multirow{9}{*}{Mnist}     & \multirow{9}{*}{\rotatebox{0}{LeNet-5}}   & R-V-W \cite{charan2020accurate}  & 16                                                    & \cmark & 0.1                                                  & 0                                                             & \xmark                                                       & Multi-Write                                              & 99.51                                                        & \textbf{99.24}                                                     & 0.27                                                  \\
                           &                            & AIIR \cite{charan2020accurate}   & 16                                                    & \cmark & 0.1                                                  & 0                                                             & $\sim$15\% A. ↑                                           & Remapping                                                & 99.51                                                        & 99.13                                                     & 0.38                                                  \\
                           &                            & KD \cite{charan2020accurate}    & 16                                                    & \cmark & 0.1                                                  & 0                                                             & \xmark                                                       & \xmark                                                    & 99.51                                                        & 97.43                                                     & 2.08                                                  \\
                           &                            & Co.Net \cite{eldebiky2022correctnet} & NA                                                    & \xmark & 0.5                                                  & 0                                                             & \xmark                                                       & $\sim$5\% Weights ↑                                      & 98.79                                                        & 97.47                                                     & 1.32                                                  \\
                           &                            & CTSF  \cite{huang2022rescuing}  & 8                                                     & \xmark & 0.5                                                  & 0                                                             & $\sim$10\% A. ↑                                         & \xmark                                                    & 98.91                                                        & 98.67                                                     & \textbf{0.24}                                                  \\
                           &                            & \ours    & 8                                                     & \cmark & 0.1                                                  & 0                                                             & \xmark                                                       & \xmark                                                    & 99.21                                                        & 99.15                                                     &\textbf{ 0.06}                                                  \\
                           &                            & \ours    & 8                                                     & \cmark & 0.5                                                  & 0                                                             & \xmark                                                       & \xmark                                                    & 99.21                                                        & \textbf{98.90}                                                     & 0.31                                                  \\
                           &                            & \ours    & 8                                                     & \cmark & 0.1                                                  & \textbf{94.89}                                                         & \xmark                                                       & \xmark                                                    & 99.21                                                        & 98.60                                                     & 0.61                                                  \\
                           &                            & \ours    & 8                                                     & \cmark & 0.5                                                  & \textbf{94.89}                                                        & \xmark                                                       & \xmark                                                    & 99.21                                                        & 97.31                                                     & 1.90                                                  \\ \hline \hline
\multirow{13}{*}{Cifar-10} & \multirow{10}{*}{\rotatebox{0}{VGG-16}}    & R-V-W \cite{charan2020accurate} & 16                                                    & \cmark & 0.1                                                  & 0                                                             & \xmark                                                       & Multi-Write                                              & 93.35                                                        & 88.43                                                     & 4.92                                                  \\
                           &                            & AIIR \cite{charan2020accurate}  & 16                                                    & \cmark & 0.1                                                  & 0                                                             & $\sim$15\% A. ↑                                           & Remapping                                                & 93.35                                                        & \textbf{91.86}                                                     & \textbf{1.49}                                                  \\
                           &                            & KD \cite{charan2020accurate}    & 16                                                    & \cmark & 0.1                                                  & 0                                                             & \xmark                                                       & \xmark                                                    & 93.35                                                        & 87.13                                                     & 6.22                                                  \\
                           &                            & Co.Net \cite{eldebiky2022correctnet} & NA                                                    & \xmark & 0.5                                                  & 0                                                             & \xmark                                                       & $\sim$0.58\% Weights ↑                                   & 93.20                                                        & \textbf{91.29  }                                                   & \textbf{1.91}                                                  \\
                           &                            & CTSF  \cite{huang2022rescuing}  & 8                                                     & \xmark & 0.5                                                  & 0                                                             & $\sim$10\% A. ↑                                         & \xmark                                                    & 93.21                                                        & 91.06                                                     & 2.15                                                  \\
                           &                            & VACTSF  \cite{huang2022rescuing}  & 8                                                     & \xmark & 0.5                                                  & 9.03                                                        & $\sim$10\% A. ↑                                          & \xmark                                                    & 93.21                                                        & 86.75                                                     & 6.46                                                  \\ &                            & DRD  \cite{long2019design}  & 8                                                     & \xmark & 0.1                                                 & 0                                                        &  \xmark                                          & \xmark                                                    & 93.28                                                        & 92.76                                                     & 0.52                                                  \\
                           &                            & \ours    & 8                                                     & \cmark & 0.1                                                  & \textbf{88.25}                                                       & \xmark                                                       & \xmark                                                    & 93.28                                                        & 91.09                                                     & 2.19                                                  \\
                           &                            & \ours    & 8                                                     & \xmark & 0.1                                                  & \textbf{88.25}                                                       & \xmark                                                       & \xmark                                                    & 93.28                                                        & \textbf{93.20}                                                     & \textbf{0.08 }                                                 \\
                           &                            & \ours    & 8                                                     & \xmark & 0.5                                                  & \textbf{88.25}                                                       & \xmark                                                       & \xmark                                                    & 93.28                                                        & 90.31                                                     & 2.97                                                  \\ \cline{2-12} 
                           & \multirow{3}{*}{\rotatebox{0}{ResNet-56}} & \ours    & 8                                                     & \cmark & 0.1                                                  & \textbf{51.36}                                                       & \xmark                                                       & \xmark                                                    & 93.34                                                        & \textbf{92.08}                                                     &\textbf{ 1.26 }                                                 \\
                           &                            & \ours    & 8                                                     & \xmark & 0.1                                                  & \textbf{51.36}                                                       & \xmark                                                       & \xmark                                                    & 93.34                                                        &\textbf{ 93.18}                                                     &\textbf{ 0.16  }                                                \\
                           &                            & \ours    & 8                                                     & \xmark & 0.5                                                  & \textbf{51.36}                                                       & \xmark                                                       & \xmark                                                    & 93.34                                                        & \textbf{91.88}                                                     & \textbf{1.46 }                                                 \\ \hline
\end{tabular}
\end{table}


\subsection{Evaluation of Whole Comprehensive Learning Framework}

\subsubsection{Evaluation of the impact from pruning \& crossbar height}

To investigate the impact of pruning and crossbar height on the accuracy of DNN models when deployed on non-ideal ReRAM devices, we conducted an analysis of the changing trend of LeNet-5 accuracy on the Mnist dataset under different degrees of non-ideality  (by changing the $\epsilon$ in Eq.(\ref{eq:1})), as shown in Fig.~\ref{fig:acc-h-sigma}. The pruning rate is the same as that presented in Table \ref{tab:pq}, where 94.89\% of the weights are removed. As the non-ideality of ReRAM cells is random, we performed 20 inferences for each trained model, and the line in Fig.~\ref{fig:acc-h-sigma} represents the average of 20 inferences, while the shaded part denotes the range of the accuracy. From this figure, we can find that without pruning, the accuracy of the network does not decrease significantly with the device's non-ideality increasing under non-ideality adaptation (NIA). Since Mnist is a small dataset and LeNet-5 has a high redundancy in the Mnist dataset, NIA training only needs to train most of the weights close to zero. As many zeros are equivalent to "reducing the crossbar height", the accuracy will not decrease much. Therefore, the unpruned redundant model is more tolerant of the device's non-ideality. In contrast, under network pruning, we found that NIA can increase the accuracy of the network on non-ideal devices. The accuracy without NIA dropped very quickly, and the average accuracy is only 90.54\% with $\epsilon = 0.8$. However, under the same crossbar height, the accuracy of NIA is 96.54\%. Also the height of the crossbar has a significant influence on the accuracy of the model. The larger the height, the lower the accuracy under the same non-ideality degree ($\epsilon$). This finding supports the conclusion derived in Section \ref{subsection:variation}.

\subsubsection{Evaluation of final compact \& reliable models for IMP}

Table \ref{tab:all} presents a comparison of our proposed method with the state-of-the-art non-ideality processing methods, including R-V-W \cite{charan2020accurate}, AIIR \cite{charan2020accurate}, KD \cite{charan2020accurate}, CorrectNet (Co.Net) \cite{eldebiky2022correctnet}, CTSF \cite{huang2022rescuing}, VACTSF \cite{huang2022rescuing}, and DRD \cite{long2019design}. We follow these methods to demonstrate the performance of our method on Mnist and Cifar-10 datasets, and the optimal values for each metric are indicated in \textbf{boldface}. To ensure a fair comparison of accuracy, we also report the quantization bit number (Quan. level), whether the methods consider SAF, the degree of variation (Varia. ($\epsilon$)), and whether the models are compressed. Additionally, we report the overhead introduced by each method. To compare with each method,  we evaluate our method under two degrees of device variation ($0.1$ and $0.5$) and with and without SAF. From Table \ref{tab:all}, we can conclude that, on the Mnist dataset and without model compression, our method achieves similar inference accuracy compared to the other methods, even though our method is with fewer quantization bits or considering SAF, and without incurring overhead. With compression in our method, it only introduces a slight accuracy drop, especially for $\epsilon=0.1$. Under the VGG-16 model in the Cifar-10 dataset,  compared to R-V-W \cite{charan2020accurate}, AIIR \cite{charan2020accurate}, and KD \cite{charan2020accurate}, our method achieves similar accuracy even with model compression, smaller quantization bit, and no introducing overhead. Compared to DRD \cite{long2019design}, our method considers the runtime characteristics of the crossbar to improve the accuracy  significantly. Compared to VACTSF \cite{huang2022rescuing}, which is CTSF \cite{huang2022rescuing} with pruning, our method achieves a much larger compression ratio and higher precision without introducing overhead.  Furthermore, to demonstrate the applicability of our method to different model architectures, we also show its results on the ResNet-56 model. And our method can still achieve good performance. Overall, our method achieves compact and reliable models with a slight loss of accuracy (i.e., $2.19\%$ and $1.26\%$ for VGG-16 and ResNet-56 on the Cifar-10 dataset, respectively), which indicates it is a promising approach for deploying DNN on ReRAM IMP accelerators.

\section{Conclusion}
\label{section:conclusion}

This work presented a comprehensive learning approach, named \ours, for achieving compact and reliable IMP acceleration. First, a multi-grained crossbar-aligned pruning method was introduced, which included kernel-group pruning and crossbar pruning to save crossbars without additional hardware. Second, a simple yet efficient integer-only quantization scheme was proposed to avoid using FP multipliers in IMP devices. Third, a runtime-aware non-ideality adaptation scheme was presented, using a realistic crossbar-based runtime simulator to learn reliable DNN models  for IMP devices  without any overhead. Finally, a novel learning framework was designed to complete these three schemes and co-optimize them, reducing training overhead and enhancing accuracy. To further improve accuracy, this learning framework optimized the model with a dynamic zero-recovery procedure for better architecture and weights. The evaluation results showed that, compared to the original model, \ours achieved a $122.38\times$ reduction in computing power and a $19.51\times$ savings in computing area on average. Moreover, \ours obtained totally integer-only, pruned, and reliable VGG-16 and ResNet-56 models for the Cifar-10 dataset on ReRAM-based IMP devices with only $2.19\%$ and $1.26\%$ accuracy drops, respectively, without any additional hardware overhead.

\section*{Acknowledgment}
This work is partially supported under the RIE2020 Industry Alignment Fund – Industry Collaboration Projects (IAF-ICP) Funding Initiative, as well as cash and in-kind contribution from the industry partner, HP Inc., through the HP-NTU Digital Manufacturing Corporate Lab (I1801E0028), and partially supported by the Ministry of Education, Singapore, under its Academic Research Fund Tier 2 (MOE2019-T2-1-071), and Nanyang Technological University, Singapore, under its NAP (M4082282/04INS000515C130).

\bibliographystyle{ACM-Reference-Format}
\bibliography{sample-base}





\end{document}